\documentclass[10pt,aps,physrev,twocolumn,superscriptaddress,floatfix,longbibliography]{revtex4-2}

\usepackage{graphicx}
\usepackage{subfigure}
\usepackage{dcolumn}
\usepackage{bm}
\usepackage{amsmath}
\usepackage{xcolor}
\usepackage{ulem} 

\begin{document}

\title{Low-energy structure and topology of the two-band Hubbard-Kanamori model}

\author{Nayara G. Gusmão}\email{Contact author: nayaragusmao@ufmg.br}
\affiliation{Departamento de Física, Universidade Federal de Minas Gerais, C. P. 702, 30123-970, Belo Horizonte, MG877}
\affiliation{Instituto de Nanociencia y Nanotecnología CNEA-CONICET, Centro Atómico Bariloche and Instituto Balseiro, Bariloche, Argentina}

\author{ Germán Blesio}
\affiliation{Instituto de F\'{\i}sica Rosario (CONICET) and Facultad de Ciencias Exactas, Ingeniería y Agrimensura (UNR)}

\author{Armando Aligia}
\affiliation{Instituto de Nanociencia y Nanotecnología CNEA-CONICET, Centro Atómico Bariloche and Instituto Balseiro, Bariloche, Argentina}

\author{Walber H. Brito}
\affiliation{Departamento de Física, Universidade Federal de Minas Gerais, C. P. 702, 30123-970, Belo Horizonte, MG877}

\author{Maria C. O. Aguiar}
\affiliation{Departamento de Física, Universidade Federal de Minas Gerais, C. P. 702, 30123-970, Belo Horizonte, MG877}

\author{Karen Hallberg}
\affiliation{Instituto de Nanociencia y Nanotecnología CNEA-CONICET, Centro Atómico Bariloche and Instituto Balseiro, Bariloche, Argentina}

\date{\today}

\begin{abstract}
We investigate the Mott transition in a two-band Hubbard-Kanamori model using Dynamical Mean-Field Theory (DMFT) with the Density Matrix Renormalization Group (DMRG) and the Numerical Renormalization Group (NRG) as impurity solvers. The study focuses on the case where the intraorbital and interorbital Coulomb interactions are equal ($U = U_2$) and the Hund's coupling is absent ($J = 0$). Our spectral analysis for this case confirms the absence of an orbital-selective Mott transition (OSMT) even in systems with very distinct bandwidths ($t_1$ and $t_2$ parameters for wide and narrow band, respectively) indicating a simultaneous Mott transition occuring in both bands. Notably, the NRG results show the emergence of a pseudo-gap-like feature and a central peak in the narrow band, whose characteristics depend on the hopping parameter $t_2$. These spectral features may act as precursors to OSMT in more realistic systems with finite Hund’s coupling ($J > 0$). Furthermore, in the Mott insulating phase, the self-energies for both bands diverge, indicating that the Mott transition represents a topological phase transition for both bands. Our results underscore the importance of accurate impurity solvers in capturing the density of states and detailed spectral structures.
\end{abstract}

\maketitle

\section{\label{sec:level1}Introduction }

Strongly correlated materials with multi-orbital electronic states have attracted great scientific interest due to the intricate, and hard to describe, interplay between the electronic interactions, crystal field splittings, and spin-orbit coupling~\cite{Aichhorn_2017}.
In these systems, the strength of electronic correlations are ruled by the interplay of Hubbard interactions and Hund's coupling. These systems present emergent phenomena not found in single-orbital systems, such as orbital-selective correlations, which is a universal feature across the iron chalcogenides~\cite{Yi2015}, for instance. In some cases, these orbital dependent correlations can induce an orbital-selective Mott transition (OSMT), which were early used to explain the existence of local moments within a metallic state in the ruthenates~\cite{anisimov2002orbital}. The OSMT is characterized by a Mott metal-insulator transition occurring in one orbital while others remain metallic, highlighting the role of orbital interactions in the formation of insulating and metallic states.

Iron-based superconductors (FeSCs) are examples of compounds for which there are both experimental and theoretical evidences of an OSMT~\cite{patel2019fingerprints, yu2021orbital}. This underscores the critical role of orbital-selective behavior in understanding the superconducting properties of FeSCs~\cite{huang2022correlation}. Similarly, ruthenates, such as Ca$_{2-x}$Sr$_2$RuO$_4$ (CSRO), exhibit orbital differentiation driven by the interplay of moderate electronic correlations and strong Hund coupling~\cite{kim2022signature, georges2013strong}. While FeSCs, with their $3d$ orbitals, exhibit stronger electronic correlations and non-conventional superconductivity at high critical temperatures ($T_c$), mediated by spin and orbital fluctuations, ruthenates, characterized by more delocalized $4d$ orbitals, also display non-conventional superconductivity alongside complex magnetic behavior. The coexistence of localized and itinerant electronic states highlights the central role of OSMT in these emergent phenomena, emphasizing the need to refine theoretical frameworks to elucidate its underlying physics.

In this study, we aim to investigate the spectral properties of the two-band Hubbard-Kanamori model, which presents the so-called narrow and wide bands, according to their different bandwidths, $t_2$ and $t_1$, respectively. We are particularly interested in the region near the phase transition where the narrow band begins to exhibit an in-pseudogap structure with quasi-particle states, indicated by region II in Fig.~\ref{fig:ModeloHK}(a), as we will detail below. To analyze this complex system, we employed the dynamical mean-field theory (DMFT)~\cite{georges1996dynamical}, combined with the density matrix renormalization group (DMRG)~\cite{Garcia2004,nunez2018solving} and with the numerical renormalization group (NRG)\cite{bulla2008numerical}, both of which serve as impurity solvers within the DMFT framework.

NRG gives an excellent description of spectral properties at low energies, close to the Fermi level, while DMRG provides high-precision results at higher energies, enabling a detailed examination of Hubbard bands and the gap in the insulating regime. We applied these methods to investigate the occurrence of OSMT in the symmetric case, where the intra- ($U$) and interorbital ($U_2$) Coulomb interactions are equal and the Hund's coupling ($J$) is neglected ($U=U_2$ and $J=0$). 

In previous work\cite{nunez2018emergent} it was shown that, for these parameters, there are well-defined quasiparticle states with a high weight in holon-doublon inter-orbital bound states (Fig.~\ref{fig:ModeloHK} (b)) located at the Fermi energy $\omega=0$. As was concluded in that work, these states are responsible for the absence of an OSMT at any small value of $t_2/t_1$ with $t_2<t_1$.

Building on this foundation, the present work goes beyond Ref.~\cite{nunez2018emergent} by providing a detailed characterization of the low-energy spectral features associated with this bound state formation.

In this work we observe an in-pseudogap spectral structure emerging exclusively in the narrow band (NB) near the Mott transition (region II in Fig.~\ref{fig:ModeloHK}(a)) while the wide band (WB)  exhibits a pronounced Kondo peak. This structure is characterized by finite DOS near the Fermi level, whose shape is influenced by the bandwidth ratio $t_2/t_1$. A region with a very low density of states in the narrow band, which we will call a pseudogap, is observed, which significantly increases in scale for very small values of $t_2/t_1$, while still maintaining a narrow central peak. 
By exploring the physics of the Hubbard bands for different values of $U$, we confirm that the Mott transition occurs simultaneously in both orbitals (region III in Fig. 1b) in agreement with studies that indicate the absence of an OSMT for the case~$J=0$~\cite{liebsch2003absence,koga2004orbital,nunez2018emergent}. In summary, in this paper, we resolve the fine structure of the metallic phase near the transition, quantify the evolution of the narrow band pseudogap, and identify signatures of the Mott criticality with improved
numerical precision.

The paper is organized as follows: Section \ref{model_methods} introduces the Hubbard-Kanamori Hamiltonian and outlines the theoretical methodologies employed in the analysis, including DMFT, with DMRG and NRG as complementary impurity-solver methods. Section \ref{results_discussion} explores the spectral properties of the model, emphasizing the behavior of the density of states (DOS) and the absence of an OSMT, even for bands with very different widths. Section \ref{topo} discusses the topological aspects of the metal-insulator transitions. Finally, the conclusions are summarized in Section \ref{conclusion}. 

\section{\label{model_methods}Model and Methods}

\begin{figure}[h]
    \centering
    \subfigure[]{\includegraphics[width=\linewidth]{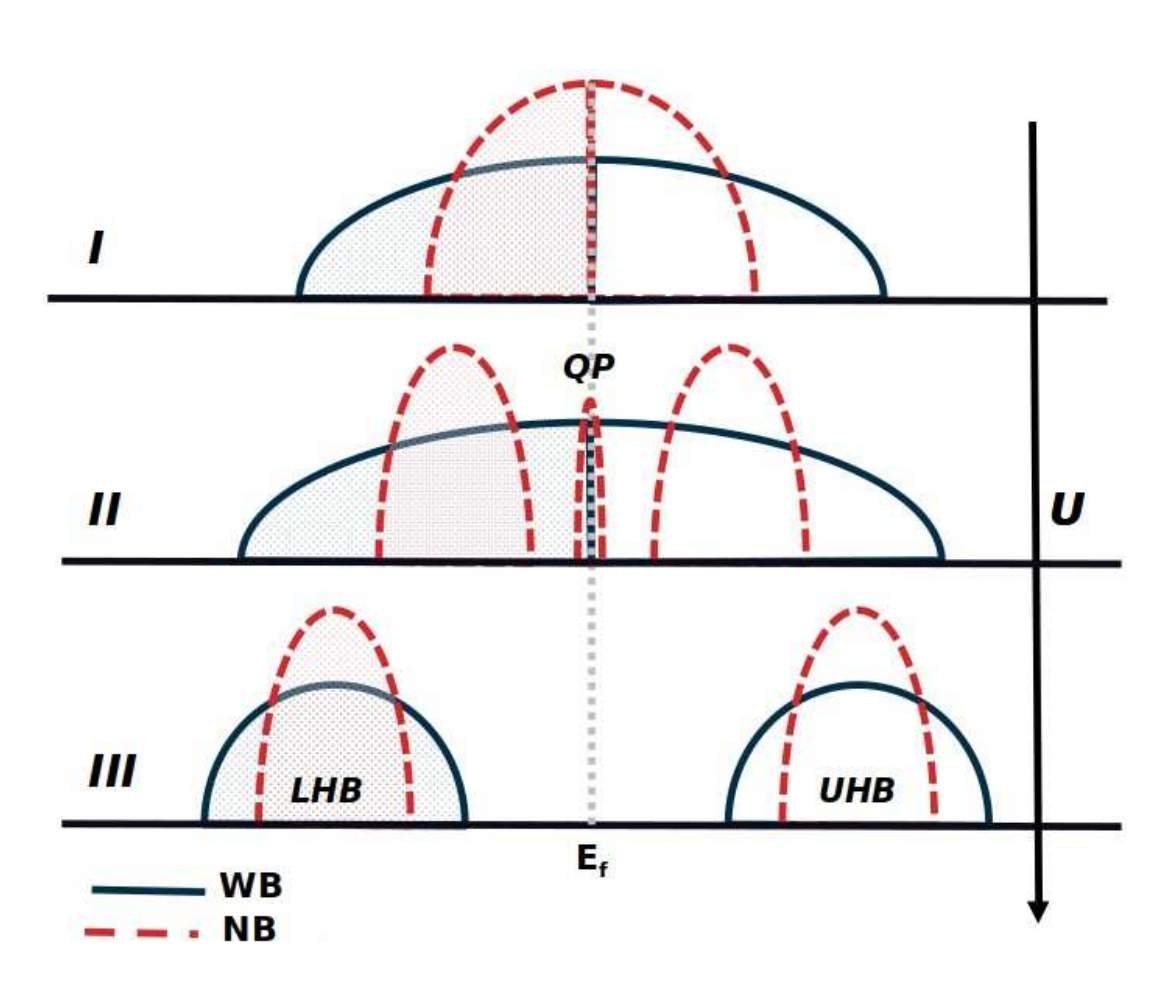}}
    \subfigure[]{\includegraphics[width=\linewidth]{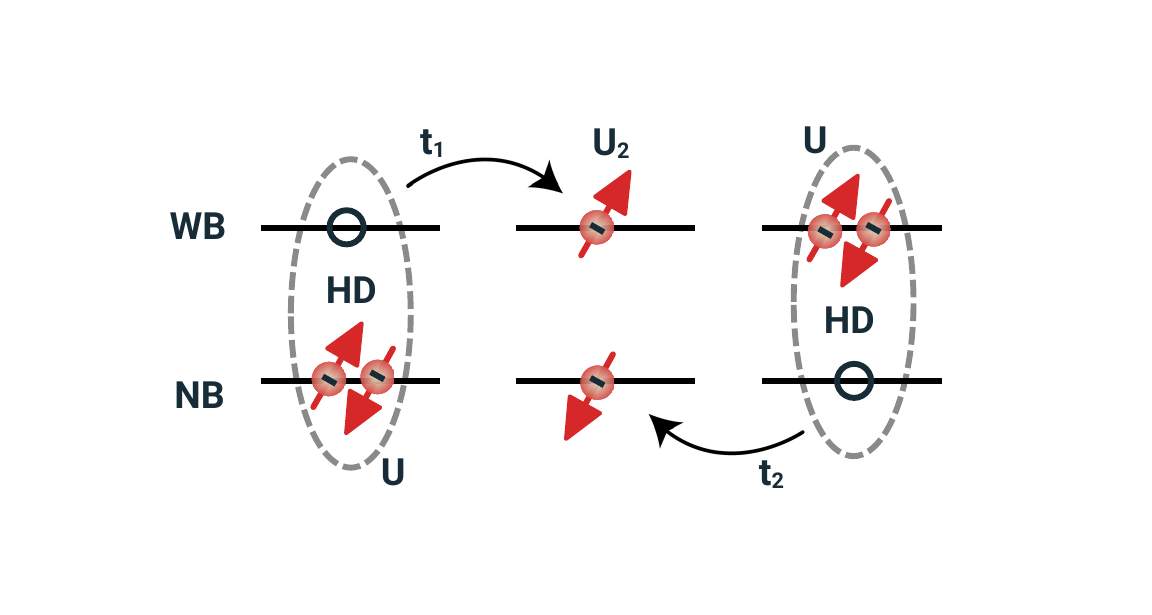}}
    \caption{(a) Illustration of the orbital-selective behaviour in a two-orbital model, depicting three distinct situations: I) At small $U$, both bands are metallic; II) At intermediate $U$, the NB presents an in-gap structure with quasiparticle (QP) states; III) At large $U$, both bands become insulating, featuring the Lower Hubbard Band (LHB) and Upper Hubbard Band (UHB). If an orbital-selective Mott transition (OSMT) exists, it would occur between (II) and (III). (b) Schematic representation of the Hubbard-Kanamori Hamiltonian for two orbitals, illustrating the formation of the holon-doublon (HD) state, characterized by an empty site in one orbital combined with a doubly occupied site in another orbital.  Additionally, the diagram depicts the electronic configurations corresponding to different interaction terms in the model: the intraorbital Coulomb interaction $U$, the interorbital Coulomb interaction $U_2$. The narrow band (NB) is associated with a smaller bandwidth, while the wide band (WB) has a larger bandwidth.}
    \label{fig:ModeloHK}
\end{figure}

We consider the Hubbard-Kanamori model with two orbitals~\cite{georges2013strong}:

\begin{equation}\label{eq:kanamori-hubbard}
H=\sum_{\langle i j\rangle I \sigma} t_I c_{i I \sigma}^{\dagger} c_{j I \sigma}+\sum_i \hat{V}_i,
\end{equation}
where $c_{i I \sigma}^{\dagger}$ ($c_{i \bar{I} \sigma}$) represents an electron creation (annihilation) operator with spin $\sigma$ at site indexed by $i$ and where $I=1,2$ denotes orbital indices. The nearest neighbor hopping parameters are $t_1 \geq t_2$, for the wide and narrow bands respectively. The term $\hat{V}_i$ describes the electronic interaction, expressed as follows:

\begin{equation}\label{eq:intHK2b}
\begin{aligned}
\hat{V}_i=U & \sum_I n_{i I \uparrow} n_{i I \downarrow}+\sum_{\sigma \sigma^{\prime}}U_{2} n_{i 1 \sigma} n_{i 2 \sigma^{\prime}},
\end{aligned}
\end{equation}
where $U~(U_2)$ is the intra (inter)~orbital Coulomb repulsion between electrons and $n_{i } = \sum_{I\sigma}c^{\dagger}_{i I\sigma}c^{}_{i I\sigma}$ is the number operator for each orbital. Figure~\ref{fig:ModeloHK}(b) schematically illustrates the interactions in the two-band case.

In this study, we solve the Hamiltonian at half filling. We consider a semicircular density-of-states (DOS) in the non-interacting, single-band case, which corresponds to a Bethe lattice. At this point, the model has electron-hole symmetry for the Bethe or bipartite lattices. As mentioned in the Introduction, we combine DMFT~\cite{georges1996dynamical} with two impurity solvers: NRG~\cite{bulla2008numerical}  method for analyzing the spectral density near the Fermi level and DMRG~\cite{Garcia2004,nunez2018solving} method for examining phase transitions at large $U$ values.
For both methods, the calculations were performed on the real axis energy, providing insights into the system's behavior across different regimes.

To explore the competition between the metallic and Mott insulator phases in the two-orbital model, we calculate the Green's function and the self-energy. By applying DMFT in a multiorbital context~\cite{hallberg2015state}, we map the original lattice problem onto an impurity problem, which must reproduce the same local Green's function and self-energy as the original lattice model. This mapping allows us to study the local correlations using a solvable Anderson Impurity Model (AIM).The impurity Hamiltonian is given by
\begin{equation}
H_{\text{imp}} = \hat{h}_0^0 + \hat{V}_0 + H_b,
\end{equation}
where $\hat{h}_0^0$ is the local non-interacting term including onsite energies and hopping within the impurity site, and $\hat{V}_0$ is the local Kanamori interaction responsible for electron correlations at the impurity site. The bath Hamiltonian is
\begin{equation}
H_b = \sum_{I,J,q,\sigma} \lambda_{IJ}^q\, b_{Iq\sigma}^\dagger b_{Jq\sigma} + \sum_{I,J,q,\sigma} \upsilon_{IJ}^q \left( b_{Iq\sigma}^\dagger c_{0J\sigma} + \text{H.c.} \right),
\end{equation}
where $b_{Iq\sigma}^\dagger$ creates an electron in bath site $q$, orbital $I$, and spin $\sigma$. The coefficients $\lambda_{IJ}^q$ are real and symmetric, defining bath energies, while $\upsilon_{IJ}^q$ are symmetric parameters that describe the hybridization between the bath and the impurity site~\cite{nunez2018solving}.

We define the hybridization function $\Delta_I(\omega)$, which determines the non-interacting impurity Green’s function $g_{0I}(\omega) = [\omega + \mu - \Delta_I(\omega)]^{-1}$. Using DMRG or NRG, we solve the impurity problem to obtain $G^{I}_{imp}(\omega)$ and extract the self-energy $\Sigma^{I}(\omega) = g^{-1}_{0I}(\omega) - [G^{I}_{imp}(\omega)]^{-1}$. The lattice Green’s function $G^{I}_{latt}(\omega)$ is computed via DMFT self-consistency, requiring $G^{I}_{latt}(\omega) = G^{I}_{imp}(\omega)$. This condition updates $g_{0I}(\omega)$ and $\Delta_I(\omega)$ iteratively until convergence. The orbital-resolved density of states is given by $\rho^{I}(\omega) = -\frac{1}{\pi} \mathrm{Im}G^{I}_{imp}(\omega)$.

Importantly, since the model does not include interorbital hopping, the orbital index is conserved, and the self-energy remains diagonal in orbital space. This follows from a pseudospin symmetry of the Hamiltonian, which commutes with $n_{\text{wide}} - n_{\text{narrow}}$, and ensures that the impurity mapping remains valid and decoupled for all values of $t_2$.

\section{Results and discussion}
\label{results_discussion}

Our study focuses on the coexistence of itinerant and localized electrons in degenerate orbitals, characterized by distinct hopping parameters ($t_1 \geq t_2$) and no hybridization. We take the  half bandwidth of the wide band, $D_1 = 2t_1$, as the energy scale, setting $t_1 = 0.5$. We also take the origin of energies at the Fermi level $\epsilon_F=0$. This framework allows us to explore the implications of different hopping parameters on the electronic properties of the system.

\subsection{Results for small $U$}

\begin{figure}[h]
    \centering
    \includegraphics[width=\linewidth]{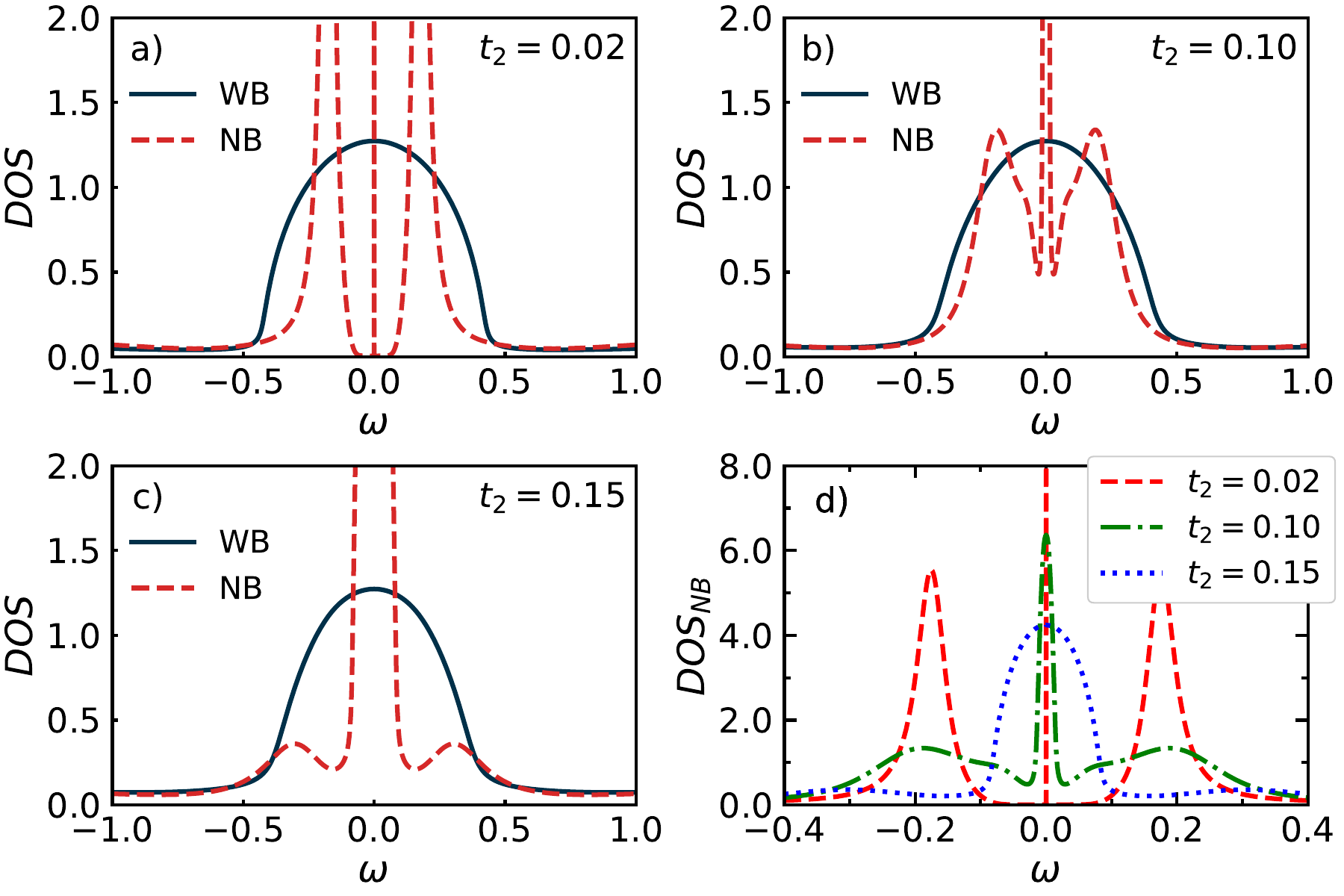}
    \caption{Electronic DOS in a metallic configuration for different values of $t_2$ with $U = U_2 = 1$ obtained using NRG as impurity solver. Panels a), b), and c) show the DOS for $t_2=0.02$, $t_2=0.1$, and $t_2=0.15$, respectively, highlighting the presence of the central peak in the NB and the broadening of the dips around it as $t_2$ decreases. Panel d) presents a zoom on the central peak of the NB, comparing the three cases and showing the influence of $t_2$ on the intensity of the DOS at $\omega = 0$. Notably, even for very small values of $t_2$, the DOS remains finite, indicating the absence of an orbital-selective Mott transition (OSMT).
    }
   \label{fig:DOSU1NRG}
\end{figure}

Our first analysis considers a metallic configuration with $U = U_2 = 1$ and varying $t_2$ values, as shown in  Fig.~\ref{fig:DOSU1NRG}.
For all $t_2$, both the NB and WB exhibit finite DOS at the Fermi level ($\omega = 0$), indicating the system’s metallic nature without evidence of an OSMT. In the NB, two side peaks form around a central peak, a structure we will refer to, in the remaining of the paper, as a pseudogap-like feature, which progressively separate from it as $t_2$ decreases. Note that this structure does not correspond to a Kondo-like peak surrounded by Hubbard bands~\cite{nunez2018emergent}. 
 
At very small $t_2$ values (e.g., $t_2 = 0.02$), the central peak in the NB becomes extremely narrow, and the pseudogap-like feature deepens. Despite these changes, the system remains metallic, even for very small $t_2$. These intricate details, particularly the narrowing of the central peak and the deepening of the “pseudogap”, are captured with high precision by the NRG method in the low-energy regime. 

In Fig.~\ref{fig:DOSU1NRG}d) we show the detail of the central peak in the NB DOS. For $U=1$, at $t_2=0.15$, the central peak exhibits a width $\sim 0.18$ and is accompanied by two broad side peaks at $\omega \sim \pm 0.3$. As $t_2$ is decreased to 0.1, the central peak narrows significantly, with a width reduced by an order of magnitude, while the side peaks intensify and shift closer to the Fermi level. For $t_2=0.02$, the width of the central peak is of the order of $3 \times 10^{-7}$. 

The intensity of the central peak in Fig.~\ref{fig:DOSU1NRG}a) is calculated as approximately 31.8, which is the expected value for the maximum possible spectral density, $\rho_I(0) = 2/(\pi t_I)$~\cite{review}, corresponding to the non-interacting case with $t_2=0.02$. Additionally, the side peaks increase in magnitude, while the intensity between these peaks and the central peak is strongly suppressed, with the minimum NB DOS approximately 0.006. Although the narrow band remains metallic for $t_2=0.02$, its contribution to metallic properties, such as specific heat, becomes negligible compared to the wide band, except at very low temperatures.

Curiously, the shape of NB DOS for small $t_2$ is qualitatively similar to that of the wide band (or the density of states for a one-band model) in the metallic side near transition to the insulating state, with a central Kondo peak and lower and upper Hubbard bands. However, the meaning of the corresponding structures for the narrow band is unclear.

\subsection{Results for large $U$}

\begin{figure}[h]
    \centering
    \includegraphics[width=\linewidth]{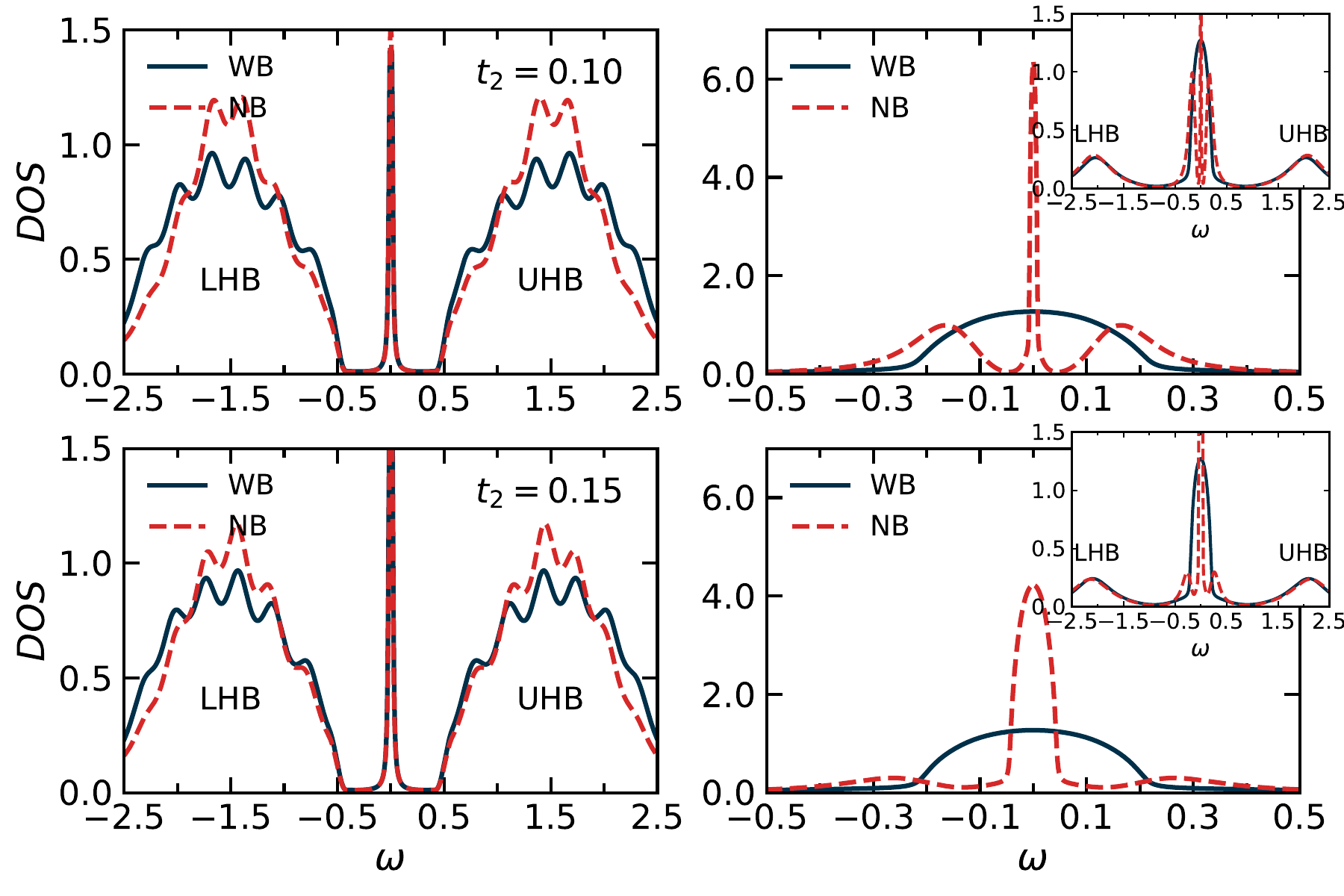}
   \caption{DOS near the Mott metal-insulator transition with $U = U_2 = 3$ and different $t_2$ values. Results were obtained using DMRG (left) and NRG (right) as the impurity solver. The insets show a broader omega region than that in the main panels. Both NB and WB exhibit signatures of the upper and lower Hubbard bands. Between these, a central Kondo-like peak emerges, with shape and spectral weight in the NB changing as $t_2$ varies.}
    \label{fig:DOSU3}
\end{figure}

Let us analyze what happens near the metal-insulator transition at $U = U_2 = 3$ for different values of $t_2$ (left panels in Fig.~\ref{fig:DOSU3}). Here, we use DMRG as the impurity solver due to its superior accuracy in capturing high-energy features at large $U$ values. While the detailed structure of the central peak cannot be fully resolved with this solver, it exhibits Kondo-like characteristics, and the Hubbard bands are well-defined and distinctly separated from the peak. The dependence on $t_2$ is evident in the relative spectral weight of the Hubbard bands and in the narrowing of the dips that surround the central peak as $t_2$ increases. 

In the right panels of Fig.~\ref{fig:DOSU3}, we use NRG with the same parameter configuration as in the left panels to investigate the behavior of the central peak in the narrow band. The results show that increasing $t_2$ the dips around the central peak are weakened.

While the main panels on the right focus on the low-energy region, the insets illustrate a broader energy range, where general agreement is observed between the NRG and DMRG results regarding the Hubbard band positions. Note that the side bands around $U/2 \sim \pm 1.5$ appear broadened in NRG. This is an artifact of NRG which uses a broadening that increases with energy. In contrast, at low energies the NRG results are more accurate than the DMRG ones, which use a constant broadening.

Due to the mentioned broadening, we must warn the reader that the height of the narrow band central peak is underestimated in the DMRG calculation. It is however correctly captured using the NRG as impurity solver. Since the self-energy $\Sigma_I(\omega)$, with $I=$ 1, 2, vanishes at the Fermi level ($\omega=0$) in the metallic phase due to electron-hole symmetry, the DOS values at the Fermi level confirm the accuracy of the NRG results when compared to the non-interacting case. Specifically, for the parameters considered in Fig.~\ref{fig:DOSU3}, the DOS at the Fermi level reaches its maximum possible values, with $\rho_1(0)=1.27$ and $\rho_2(0)=6.37$.

\begin{figure}[h]
    \centering
    \includegraphics[width=0.8\linewidth]{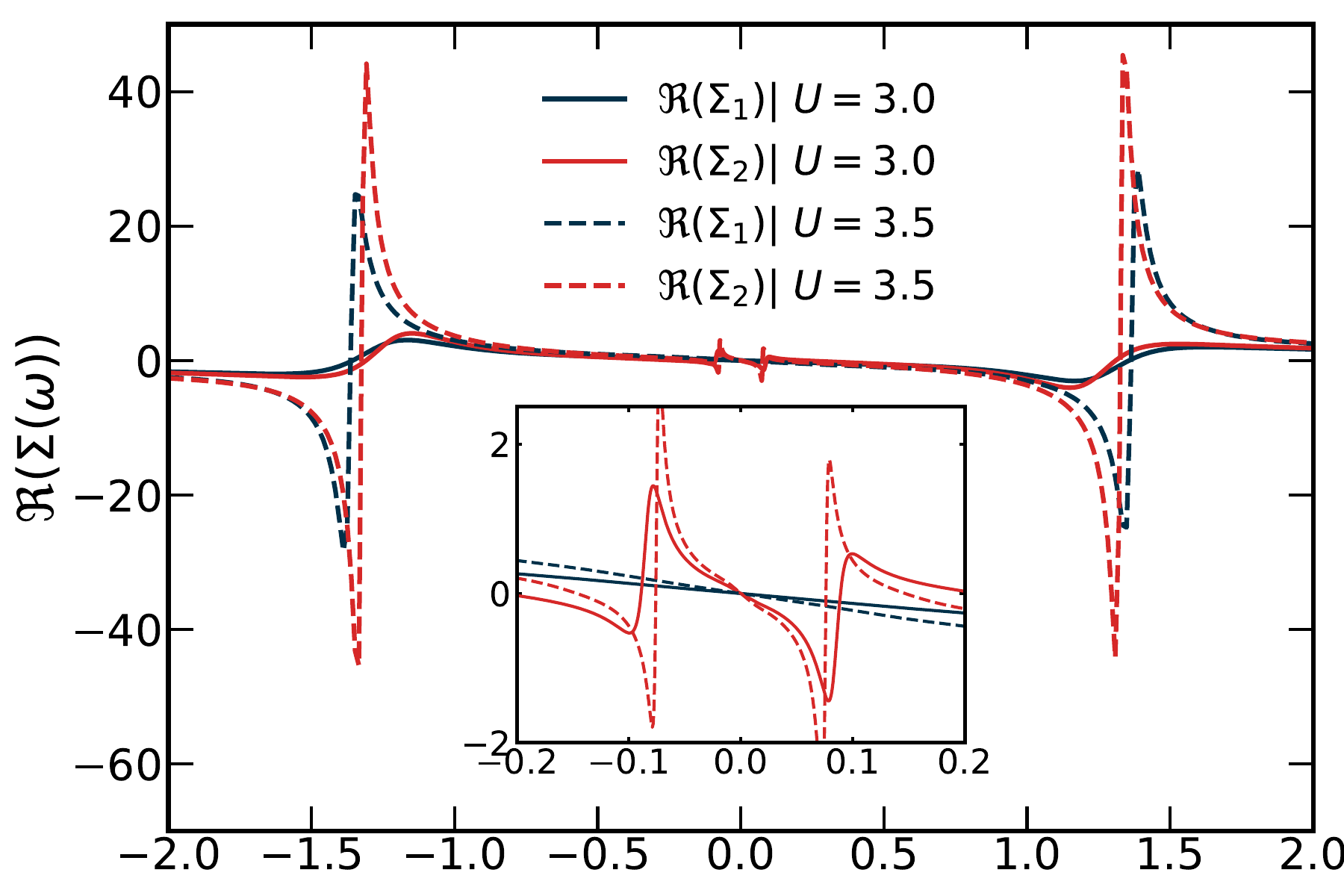}
    \includegraphics[width=0.8\linewidth]{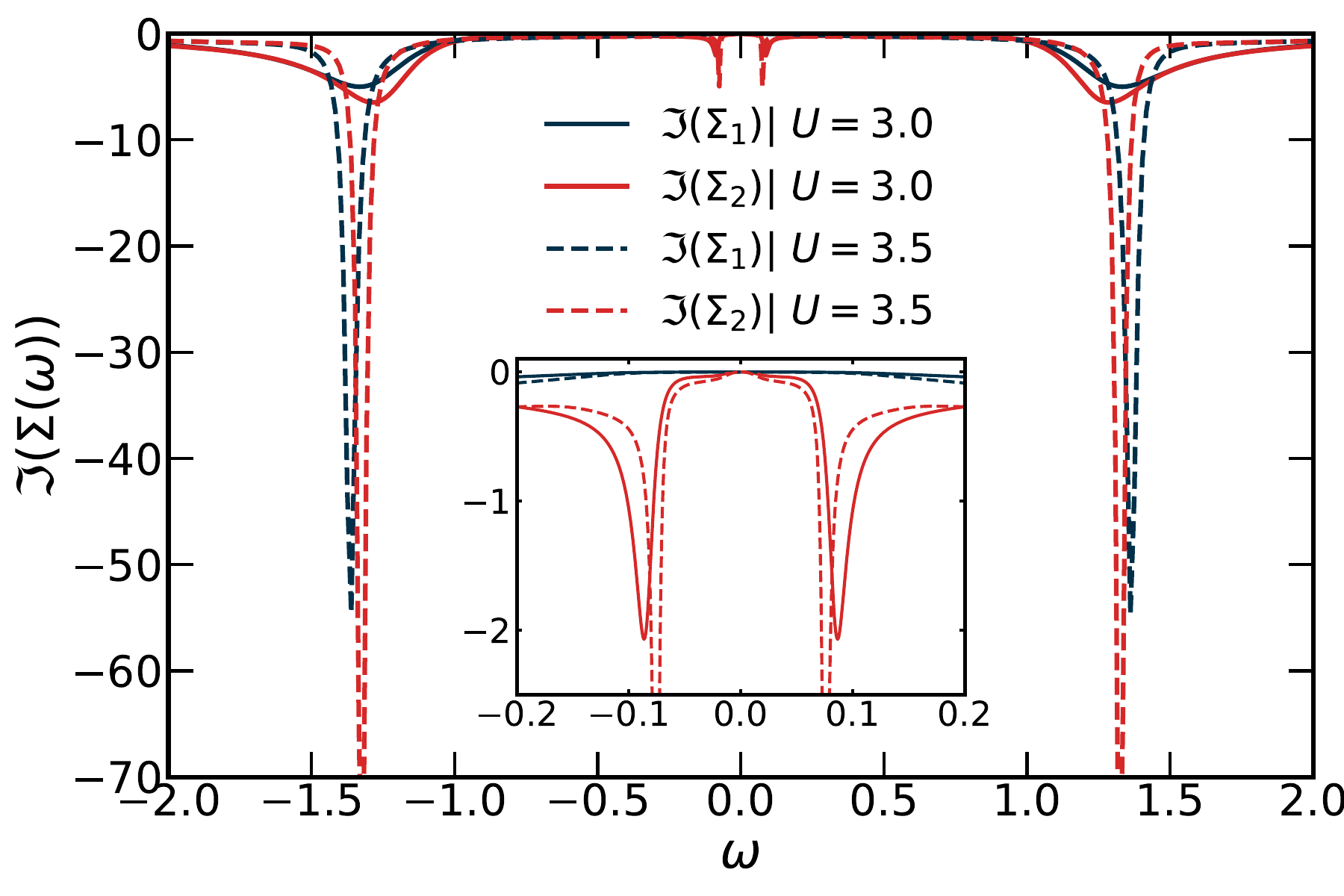}
    \caption{Real (top) and imaginary (bottom) parts of the self-energy for both orbitals, calculated with NRG as a function of energy, for $t_2 = 0.1$ and interaction strengths $U = 3$ and $U = 3.5$. The inset highlights the low-energy behavior near $\omega = 0$.}
    \label{self}
\end{figure}

According to our results in Fig. 3, both bands are metallic for $U = 3$. To confirm this conclusion and better understand it, let us analyze the corresponding self energy. In particular, the structure of $\rho_2(\omega)$ for small energy reflects the behavior of the self energy. $\Sigma_I(\omega)$ for both bands are represented in Fig.~\ref{self}. Due to electron-hole symmetry, the real  (imaginary) parts are odd (even) under a change of sign of $\omega$. The peaks in $\Im$($\Sigma_I(\omega)$) are related with dips in $\rho_I(\omega)$. In contrast, zeros in the real part of $\Sigma_I(\omega)$ are correlated with peaks in $\rho_I(\omega)$. Remarkably, the self energies of both bands are very similar for $|\omega|>0.5$. For $|\omega|<0.5$, the self energy of the broad band $\Sigma_1(\omega)$ is rather featureless, while $\Re$($\Sigma_2(\omega)$) has zeros near $\omega \sim \pm 0.18$ and at $\omega =0$ where $\rho_2(\omega)$ has peaks. For intermediate values of $\omega$, $|\Sigma_2(\omega)|$ has peaks, which correspond to the dips in $\rho_2(\omega)$. Decreasing $t_2$ these peaks in $|\Sigma_2(\omega)|$ increase in magnitude implying lower values of $\rho_2(\omega)$.

By increasing $U$ towards the Mott transition, the structures of both self-energies move towards $\omega=0$. Specifically for negative $\omega$, the upturns in the real and imaginary parts of the self energies
(near $\omega \sim -1.5$ for $\Sigma_1(\omega)$ in Fig. \ref{self}) move to smaller values of $-\omega$ and the negative values increase in magnitude and diverge at the transition. For positive $\omega$ the real parts of both self-energies change sign, as they are odd functions of $\omega$. This is similar to what is known in the one-band case \cite{sen}.

\begin{figure}[h]
    \centering
    \includegraphics[width=\linewidth]{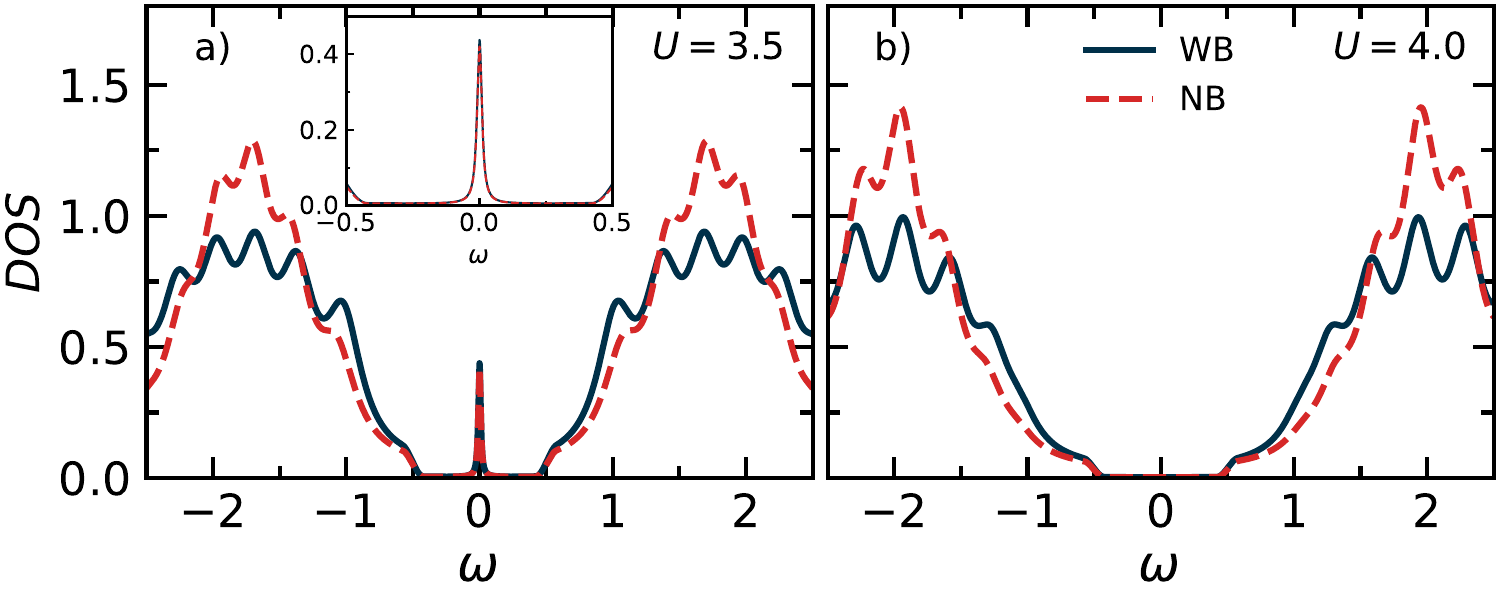}
    \caption{DOS near the metal-insulator transition with DMRG for $t_2=0.1$. We see the formation of the gap between a) $U = 3.5$ and b) $U = 4$ for both bands, confirming the absence of an OSMT. The inset in a) highligths the central peak for $U=3.5$. 
    }
    \label{fig:DOSt201DMRG}
\end{figure}

Fig.~\ref{fig:DOSt201DMRG} shows DMRG results for U=3.5 and U=4.0, highlighting the gap formation between the Hubbard bands. The zoom on the low-energy region (insets) confirms that NB and the WB are metallic for $U = 3.5$ and that the transition for both band occurs within a similar $U$ range, approximately between $3.5$ and $4$ for $t_2 = 0.1$. Notably, the same $U$ range for the transition is also found for very small $t_2$, reinforcing the absence of OSMT in these cases, with both bands transitioning simultaneously.

Our results indicate the absence of an OSMT for $U=U_2$ $(J=0)$ with $t_2/t_1<<1$, in agreement with previous studies~\cite{liebsch2003absence,koga2004orbital,nunez2018emergent}, but contradicting others that suggest that any finite $t_2$ induces OSMT~\cite{ferrero2005dynamical,de2005orbital,winograd2014hybridizing,ruegg2005slave,inaba2006phase}. In our model, as $t_2$ decreases, the NB  structure undergoes modifications, with spectral weight being transferred from the central peak to two side peaks near it, separated from the former by two dips which broaden and decrease in intensity. We associate this structure to a pseudogap like one, as previously mentioned. The total spectral weight of the central peak decreases strongly with decreasing $t_2$. 
The pseudogap like structure in the NB DOS in the metallic phase is accompanied by a narrow peak at the Fermi energy. This feature is often missed by other methods. This fact and the very small spectral density in the pseudogap  can be mistaken as corresponding to an insulating phase, leading some studies to interpret it as an OSMT for this parameter configuration. In contrast to Ref.~\cite{ferrero2005dynamical}, which used a non-self-consistent NRG approach, our results come from a fully self-consistent DMFT cycle, allowing for better spectral resolution near the Fermi level. At $U=3.5$, the peak is well-defined and separated from the Hubbard bands, manifesting as a Kondo-type peak. This result was also observed earlier in Ref.~\cite{nunez2018emergent}, where the same method was used, thereby ensuring the validity of the result (now with more details). It confirms that we remain in the metallic phase for both wide and narrow bands. Although the OSMT is absent, for very small $t_2$, the metallic properties are dominated strongly by the wide band, except at very low temperatures.

In Ref.~\cite{kim2022signature}, the authors show that the Kondo peak and OSMT are intertwined, with Kondo hybridization playing a key role in the emergence of OSMT for $J>0$. They observe a similar change in bandwidth linked to the Kondo peak and OSMT, which agrees with our findings of spectral weight transfer and bandwidth modifications. Although our model corresponds to $J=0$, the Kondo-like peak suggests a similar mechanism at play. We propose that the observed structure in our results could be a precursor to the OSMT that would emerge in real materials where $J>0$.

\section{The metal-insulator transition as a topological quantum phase transition}
\label{topo}

In the metallic phase, the system satisfies the Luttinger theorem, which states that the volume enclosed by the Fermi surface is the same as the non-interacting one \cite{lu1,lu2}. This is true for both bands and spin indices, since the one-particle Green functions and self energies are diagonal in band and spin indices.

In the insulating phase of each band, naturally the Fermi surface disappears and the volume enclosed by it vanishes. Some years ago, Seki and Yunoki~\cite{seki} have shown that this jump in the volume enclosed by the Fermi surface is given by the corresponding jump in the winding number of a quantity $D(z)$, which in our case for each band and spin takes the form~\cite{zitko}

\begin{equation}
D_{I \sigma}(z)=\frac{g_{I \sigma 0}(z)}{G_{I \sigma}(z)},
\end{equation}
where $z$ extends $\omega$ to the complex plane.

For each band, the divergence of the self-energy for $\omega=0$ at the metal-insulator transition implies the appearance of a zero in $G_{I \sigma}(0)$ and therefore a pole in $D_{I \sigma}(z)$ at the origin. This implies a jump in its winding number and therefore, a topological transition.

The topological character of the transition for the one-band model has been discussed before \cite{sen}. The authors have shown using NRG that the Wilson chain takes the form of a generalized Su-Shrieffer-Hegger model with different topological properties at both sides of the transition.

Despite being unable to determine the critical value of $U$ at the transition using NRG due to technical limitations, our results support the existence of an analogous  topological Mott transition in the two-band case for $U = U_2$. Specifically, as $U$ increases within the
metallic phase, the peaks of the real and imaginary parts of the self-energies for both bands inside the pseudogap shift towards $\omega = 0$ and grow in magnitude. The transition occurs when these peaks merge
and diverge at $\omega = 0$. 
Curiously, a simultaneous transition for both channels has also been found in an Anderson model with two inequivalent channels including a configuration with an anisotropic spin 1, with experimental relevance \cite{zitko,review}.

\section{Conclusions}
\label{conclusion}
In conclusion, the analysis of the local electronic density of states of the two-orbital Hubbard-Kanamori model with different bandwidths, same intra- and inter-orbital Coulomb repulsions ($U=U_2$) and no Hund interaction ($J=0$) using {the DMFT with DMRG and NRG as impurity solvers provides a comprehensive understanding of the spectral behaviour near the Mott metal-insulator transition. Our results confirm the absence of an OSMT in this system, with the transition occurring in the same $U$ range for both the narrow and wide bands. DMRG revealed the formation of the gap between the Hubbard bands, while NRG highlighted the influence of the $t_2$ parameter on the central peak, especially in the NB, where the shape of the peak changes with $t_2$ showing a pseudogap at very low energies. This evolution mirrors the changes observed in metallic configurations, indicating that even at small $t_2$ values, the system retains metallic characteristics, as reflected in the density of states at $\omega = 0$. However, with decreasing $t_2$, the total spectral weight of this peak decreases strongly and the metallic properties become dominated by the wide band, except at very small temperatures.

The spectral density and the self energy of the narrow band for small $t_2$ resemble qualitatively the corresponding quantities for the wide band on the metallic side near the Mott transition, but in a considerable smaller energy range (one order of magnitude smaller for our choice of parameters). Both self-energies diverge on the insulating side of the transition. As a consequence, the Mott transition in both bands have a topological character and this fact naturally explains the jump in the volume of the Fermi surface.

\begin{acknowledgments}
The authors acknowledge the financial support from the Brazilian agencies CNPq (particularly Grants 402919/2021-1 and INCT-IQ 465469/2014-0), CAPES, and FAPEMIG. They are also grateful to the National Laboratory for Scientific Computing (LNCC/MCTI, Brazil) for providing HPC resources through the SDumont supercomputer, which have contributed to the research results, URL: http://sdumont.lncc.br. We thank the Coaraci Supercomputer for computer time (Fapesp grant 2019/17874-0) and the Center for Computing in Engineering and Sciences at Unicamp (Fapesp grant 2013/08293-7). AAA acknowledges financial support provided by PICT 2020A 03661 of the Agencia I+D+i, Argentina.
\end{acknowledgments}

\bibliography{references}

\appendix
{
\section{DOS for $U = 3.0$ and $U = 3.5$ at $t_2 = 0.02$ and $t_2 = 0.1$}
\label{ap:DOS_t2002Ularge}
This appendix provides additional spectral functions for $U = 3.0$ and $U = 3.5$, obtained using DMRG and NRG solvers for $t_2 = 0.02$ and $t_2 = 0.1$.

The left panels of Fig.~\ref{fig:ap1} show DMRG results for $t_2 = 0.02$. Although limited by resolution and entanglement growth, DMRG captures a finite density of states at the Fermi level for both orbitals, indicating a metallic phase. The central peak structure changes with increasing $U$, suggesting the emergence of a low-energy feature, although its internal structure is not fully resolved.

The right panels show NRG results for $t_2 = 0.1$, where convergence is reliable and spectral resolution is higher. A pseudogap-like feature develops at low energy and becomes more pronounced as $U$ increases, with spectral weight shifting from the Fermi level to the Hubbard bands.

These results illustrate the complementarity between DMRG and NRG. While DMRG accesses very small $t_2$ values, NRG provides detailed resolution of the low-energy structure. Together, they confirm that for $J = 0$, the system remains metallic at small $t_2$, and that the metal-insulator transition occurs simultaneously in both orbitals at a common critical interaction $U_c$.

\begin{figure}
\centering
\includegraphics[width=\linewidth]{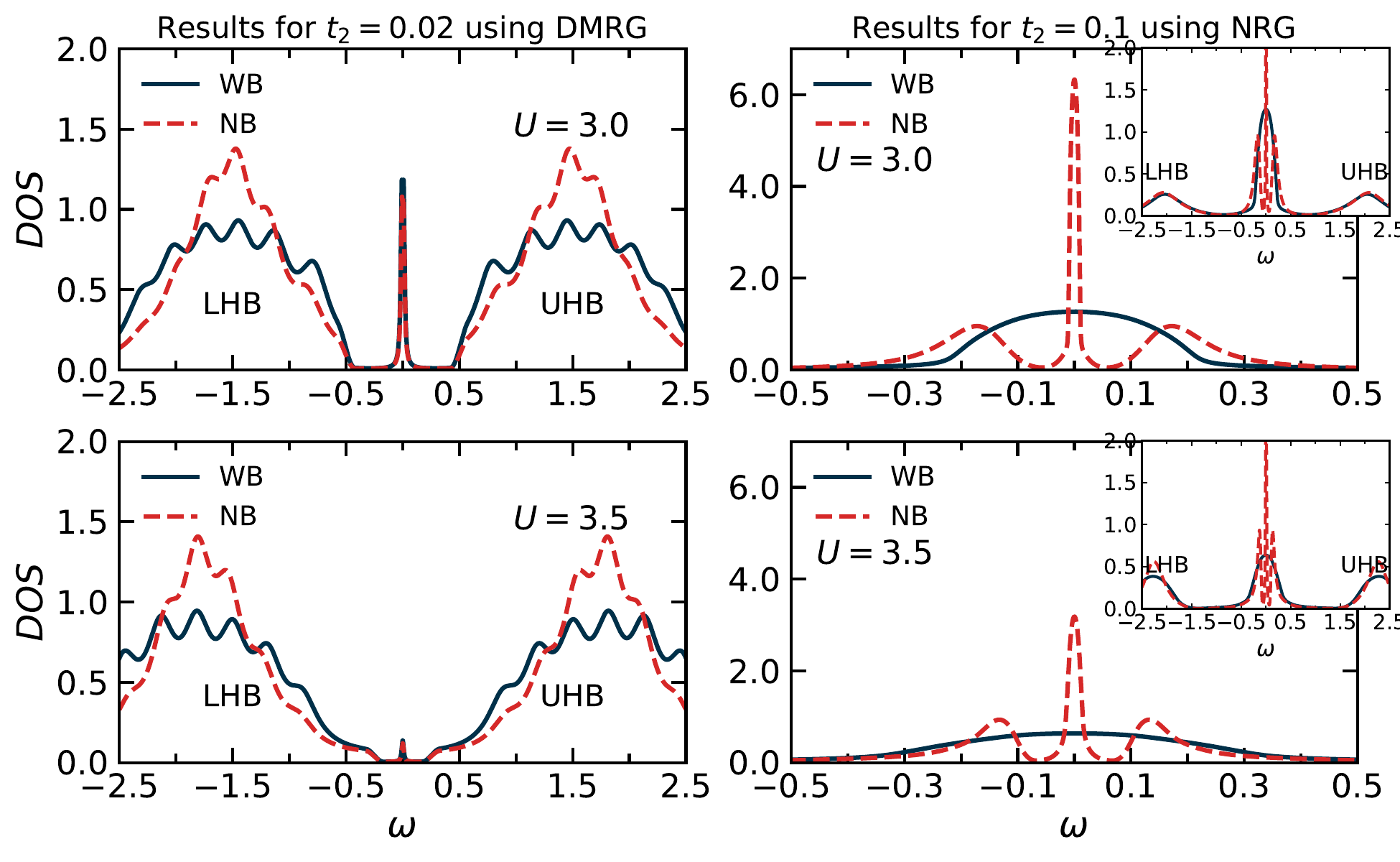}
\caption{DOS for $U = 3.0$ and $U = 3.5$. Left panels: DMRG results for $t_2 = 0.02$. Right panels: NRG results for $t_2 = 0.1$. For the NRG data, the main panels highlight the central peak structure near the Fermi level, while the insets show a broader frequency range including the lower and upper Hubbard bands.
}
\label{fig:ap1}
\end{figure}
}
\end{document}